\documentclass[aps,preprint,,showpacs,floats,epsf,epsfig,nofootinbib,12pt]{revtex4}
\usepackage{graphicx}
\usepackage{epsfig}
\usepackage[dvips]{color}

\def\beq{\begin{eqnarray}}
\def\eeq{\end{eqnarray}}
\def\non{\nonumber}
\def\lqcd{\Lambda_{\rm QCD}}
\def\vslash{\rlap{\hspace{-0.02cm}/}{v}}

\def\la{\langle}
\def\ra{\rangle}
\def\Mb{M_{\Lambda_b}}
\def\Mc{M_{\Lambda_c}}

\begin{document}

\title{ Diquarks and $\Lambda_{b}\to\Lambda_c$ weak decays }

\vspace{1cm}

\author{Hong-Wei Ke\footnote{Email: khw020056@mail.nankai.edu.cn},
        Xue-Qian Li\footnote{Email: lixq@nankai.edu.cn} and
        Zheng-Tao Wei\footnote{Email: weizt@nankai.edu.cn} }

\affiliation{ Department of Physics, Nankai University, Tianjin
 300071, China }

\vspace{12cm}

\begin{abstract}

In this work we investigate the weak  $\Lambda_{b}\to\Lambda_c$
semi-leptonic and non-leptonic decays. The light-front quark model
and diquark picture for heavy baryons are adopted to evaluate the
$\Lambda_{b}\to\Lambda_c$ transition form factors. In the heavy
quark limit we study the Isgur-Wise function. The transition form
factors are obtained in the whole physical momentum regions. The
numerical predictions on the branching ratios of non-leptonic decay
modes $\Lambda_{b}\to\Lambda_c M$ and various polarization
asymmetries are made. A comparison with other approaches is
discussed.


\end{abstract}

\maketitle

\section{Introduction}

The weak decays of $\Lambda_b$ provide valuable information of the
CKM parameter $V_{cb}$ and serve as an ideal laboratory to study
non-perturbative QCD effects in the heavy baryon system. Recently
the DELPHI collaboration reported their measurement on the slope
parameter $\rho^2$ in the Isgur-Wise function and the the braching
ratio of the exclusive semi-leptonic process
$\Lambda_b\to\Lambda_cl\bar\nu_l$ \cite{DELPHI}. This experimental
measurement re-excites great theoretical interests in
semi-leptonic decays of $\Lambda_b$ \cite{AHN,SLL,GKLLW,
HJKL,PWC,EFG}. From PDG06 \cite{PDG06}, signals of several
non-leptonic processes, such as $\Lambda_b \to \Lambda_c\pi,
\Lambda_ca_1(1260)$ have been observed. These processes are a good
probe to test the factorization hypothesis which has been
extensively explored in the B meson case \cite{Factorization}. The
forthcoming LHCb project is expected to accumulate  large samples
of the b-flavor hadrons and offer an opportunity to study in
detail the $\Lambda_b$ decays. Thus it probably is the time to
investigate the $\Lambda_b$ weak decays more systematically. In
this work, we will concentrate on the $\Lambda_b\to\Lambda_c$
semi-leptonic and non-leptonic decays.

As in the B meson decays, the key for correctly evaluation on the
rate of the semi-leptonic decays is how to properly derive the
hadronic matrix element which is parameterized by the $\Lambda_b \to
\Lambda_c$ transition form factors. There are various approaches in
the market to evaluate these form factors: the QCD sum rules
\cite{HJKL}, quark models \cite{CS,AHN,PWC,EFG}, perturbative QCD
approach \cite{SLL,GKLLW} {\it etc.}. In this work, we will study
the heavy baryon form factors in the light-front quark model. The
light-front quark model is a relativistic quark model based on the
light-front QCD \cite{light}. The basic ingredient is the hadron
light-front wave function which is explicitly Lorentz-invariant. The
hadron spin is constructed using the Melosh rotation. The
light-front approach has been widely applied to calculate various
decay constants and form factors for the meson cases
\cite{Jaus,meson2,CCH1,CCH2,HW}. Different from the case discussed
in \cite{CS} where the light-front quark model was also employed, we
adopt the diquark picture for baryons. It has been known for a long
time that two quarks in a color-antitriplet state attract each other
and may form a correlated diquark \cite{DJS}. Indeed, the diquark
picture of baryons is considered to be appropriate for the low
momentum transfer processes \cite{kroll,wilczek,yu,MQS}. In the
conventional quark model, the heavy baryon is composed of one heavy
quark and two light quarks. The light spectator quarks participate
only in the soft interactions as $\Lambda_b $ transits into
$\Lambda_c$, hence it is reasonable to employ the diquark picture
for heavy baryons where the diquark serves as a whole spectator.
Concretely, under the diquark approximation, $\Lambda_b$ and
$\Lambda_c$ are of the one-heavy-quark-one-light-diquark(ud)
structure which is analogous to the meson case and a considerable
simplification in the calculations is expected. In fact, some
non-perturbative interactions between the two light quarks can be
effectively absorbed into the constituent diquark mass.  In this
phenomenological study, we use the rate of the semi-leptonic process
$\Lambda_b\to\Lambda_c l\bar \nu_l$ which will be accurately
measured at LHCb and future ILC, to constrain the light scalar
diquark mass.

For the non-leptonic two-body decays $\Lambda_b\to\Lambda_c M$ where
$M$ denote light mesons, the amplitude is factorized to a product of
the meson decay constant and $\Lambda_b\to\Lambda_c$ transition form
factors by the factorization assumption. Because there only a
color-allowed diagram is involved, the factorization assumption is
believed to be reliable in the B meson case \cite{Factorization}.
However, the theoretical predictions on the non-leptonic two-body
decays vary by a factor of 2-3 for various models. The main
theoretical uncertainties arise from the model evaluations of the
form factors. In order to reduce model dependence and obtain a more
reliable prediction, we study the semi-leptonic decays and
non-leptonic processes simultaneously. The present experimental data
of the semi-leptonic decays (although the errors are still quite
sizable) set constraints on the phenomenological parameters in the
light-front approach. With these parameters, even though not very
accurate yet, we evaluate the $\Lambda_b\to\Lambda_c$ form factors
and make predictions on the widths of the  semi-leptonic decay
$\Lambda_b\to\Lambda_c+l\bar\nu$ and non-leptonic two-body decay
$\Lambda_b\to\Lambda_c+M$ where $M$ is a meson.

We organize our paper as follows. In section II, we formulate the
form factors for the transition $\Lambda_b\to\Lambda_c$ in the
light-front approach. We will show that in the heavy quark limit,
the resultant form factors are related to one universal Isgur-Wise
function. In section III, the formulations of the decay ratios and
the polarizations for the semi-leptonic and non-leptonic two-body
decays are given. In section IV, we present our numerical results
and all relevant input parameters are given explicitly. We then
compare our numerical results with the predictions by other
approaches. Finally, Section V is devoted to conclusion and
discussion.

\section{$\Lambda_b\to\Lambda_c$ transition form factors in light-front
 approach}

In the diquark picture, the heavy baryon $\Lambda_{b(c)}$ is
composed of one heavy quark $b(c)$ and a light diquark [ud]. In
order to form a color singlet hadron, the diquark [ud] is in a color
anti-triplet. Because$\Lambda_{b(c)}$ is at the ground state, the
diqaurk is in a $0^+$ scalar state (s=0, l=0) and the orbital
angular momentum between the diquark and the heavy quark is also
zero, i.e. $L=l=0$.

\subsection{Heavy baryon in the light-front approach}

In the light-front approach, the heavy baryon $\Lambda_Q$ with total
momentum $P$, spin $S=1/2$ and scalar diquark can be written as
\begin{eqnarray}\label{eq:lfbaryon}
 |\Lambda_Q(P,S,S_z)\rangle&=&\int\{d^3p_1\}\{d^3p_2\} \,
  2(2\pi)^3\delta^3(\tilde{P}-\tilde{p_1}-\tilde{p_2}) \nonumber\\
 &&\times\sum_{\lambda_1}\Psi^{SS_z}(\tilde{p}_1,\tilde{p}_2,\lambda_1)
  C^{\alpha}_{\beta\gamma}F^{bc}\left|\right.
  Q_{\alpha}(p_1,\lambda_1)[q_{1b}^{\beta}q_{2c}^{\gamma}](p_2)\ra,
\end{eqnarray}
where $Q$ represents $b$ or $c$, $[q_1q_2]$ represents $[ud]$,
$\lambda$ denotes helicity, $p_1,~ p_2$ are the on-mass-shell
light-front momenta defined by
\begin{equation}
 \tilde{p}=(p^+,p_{\perp}),\qquad p_\perp=(p^1,p^2),\qquad
 p^-=\frac{m^2+p_{\perp}^2}{p^+},
\end{equation}
and
\begin{eqnarray}
&&\{d^3p\}\equiv\frac{dp^+d^2 p_{\perp}}{2(2\pi)^3},\qquad
  \delta^3(\tilde{p})=\delta(p^+)\delta^2(p_{\perp}),
  \nonumber\\
&&\mid Q(p_1,\lambda_1)[q_1 q_2](p_2)\rangle=
 b^{\dagger}_{\lambda_1}(p_1)a^{\dagger}(p_2)| 0\ra,\non\\
&&[a(p'), a^{\dagger}(p)]=2(2\pi)^3\delta^3(\tilde{p}'-\tilde{p}),
  \nonumber\\
&&\{d_{\lambda'}(p'),d_{\lambda}^{\dagger}(p)\}=
  2(2\pi)^3\delta^3(\tilde{p}'-\tilde{p})\delta_{\lambda'\lambda},
\end{eqnarray}
The coefficient $C^{\alpha}_{\beta\gamma}$ is a normalized color
factor and $F^{bc}$ is a normalized flavor coefficient,
 \beq
 && C^{\alpha}_{\beta\gamma}F^{bc}C^{\alpha'}_{\beta'\gamma'}F^{b'c'}
  \la Q_{\alpha'}(p'_1,\lambda'_1)[q_{1b'}^{\beta'}q_{2c'}^{\gamma'}](p'_2)|
  Q_{\alpha}(p_1,\lambda_1)[q_{1b}^{\beta}q_{2c}^{\gamma}](p_2)\ra
  \non\\
  &&=2^2(2\pi)^6\delta^3(\tilde{p}_1'-\tilde{p}_1)\delta^3
  (\tilde{p}_2'-\tilde{p}_2)\delta_{\lambda'_1\lambda_1}.
 \eeq

In order to describe the internal motion of the constituents, one
needs to introduce intrinsic variables $(x_i, k_{i\perp})$ with
$i=1,2$ through
\begin{eqnarray}
&&p^+_1=x_1 P^+, \qquad\qquad p^+_2=x_2 P^+,
 \qquad\qquad x_1+x_2=1, \nonumber\\
&&p_{1\perp}=x_1 P_{\perp}+k_{1\perp},
  ~~~ p_{2\perp}=x_2 P_{\perp}+k_{2\perp},
  ~~~ k_{\perp}=-k_{1\perp}=k_{2\perp},
\end{eqnarray}
where $x_i$ are the light-front momentum fractions satisfing $0<x_1,
x_2<1$. The variables $(x_i, k_{i\perp})$ are independent of the
total momentum of the hadron and thus are Lorentz-invariant
variables. The invariant mass square $M_0^2$ is defined as
 \begin{eqnarray} \label{eq:Mpz}
  M_0^2=\frac{k_{1\perp}^2+m_1^2}{x_1}+
        \frac{k_{2\perp}^2+m_2^2}{x_2}.
 \end{eqnarray}
The invariant mass $M_0$ is in general different from the hadron
mass $M$ which satisfies the physical mass-shell condition
$M^2=P^2$. This is due to the fact that the baryon, heavy quark
and diquark can not be on their mass shells simultaneously. We
define the internal momenta as
 \beq
 k_i=(k_i^-,k_i^+,k_{i\bot})=(e_i-k_{iz},e_i+k_{iz},k_{i\bot})=
  (\frac{m_i^2+k_{i\bot}^2}{x_iM_0},x_iM_0,k_{i\bot}).
 \eeq
It is easy to obtain
 \begin{eqnarray}
  M_0&=&e_1+e_2, \non\\
  e_i&=&\frac{x_iM_0}{2}+\frac{m_i^2+k_{i\perp}^2}{2x_iM_0}
      =\sqrt{m_i^2+k_{i\bot}^2+k_{iz}^2},\non\\
 k_{iz}&=&\frac{x_iM_0}{2}-\frac{m_i^2+k_{i\perp}^2}{2x_iM_0}.
 \end{eqnarray}
where $e_i$ denotes the energy of the i-th constituent. The
momenta $k_{i\bot}$ and $k_{iz}$ constitute a momentum vector
$\vec k_i=(k_{i\bot}, k_{iz})$ and correspond to the components in
the transverse and $z$ directions, respectively.

The momentum-space function $\Psi^{SS_z}$ in Eq.
(\ref{eq:lfbaryon}) is expressed as
\begin{equation}
 \Psi^{SS_z}(\tilde{p}_1,\tilde{p}_2,\lambda_1)=
  \left\la\lambda_1\left|\mathcal{R}^{\dagger}_M(x_1,k_{1\perp},m_1)
   \right|s_1\right\ra
  \left\la 00;\frac{1}{2} s_1\left|\frac{1}{2}S_z\right\ra
   \phi(x,k_{\perp})\right.,
\end{equation}
where $\phi(x,k_{\perp})$  is the light-front wave function which
describes the momentum distribution of the constituents in the
bound state with $x=x_2,~k_{\perp}=k_{2\perp}$; $\left\la
00;\frac{1}{2} s_1\left|\frac{1}{2}S_z\right\ra\right.$ is the
corresponding Clebsch-Gordan coefficient with spin $s=s_z=0$ for
the scalar diquark;
$\left\la\lambda_1\left|\mathcal{R}^{\dagger}_M(x_1,k_{1\perp},m_1)
\right|s_1\right\ra$ is the well-known Melosh transformation
matrix element which transforms the the conventional spin states
in the instant form into the light-front helicity eigenstates,
 \beq
 \left\la\lambda_1\left|\mathcal{R}^{\dagger}_M(x_1,k_{1\perp},m_1)
 \right|s_1\right\ra &=& \frac{\bar u(k_1,\lambda_1)u_D(k_1,s_1)}{2m_1}
 \non\\
  &=&\frac{(m_1+x_1M_0)\delta_{\lambda_1 s_1}+i\vec{\sigma}_{\lambda_1 s_1}
  \cdot\vec k_{1\perp}\times\vec n}
  {\sqrt{(m_1+x_1M_0)^2+k_{1\perp}^2}},
 \eeq
where $u_{(D)}$ denotes a Dirac spinor in the light-front
(instant) form and $\vec n=(0,0,1)$ is a unit vector in the $z$
direction. In practice, it is more convenient to use the covariant
form for the Melosh transform matrix \cite{Jaus,CCH2}
\begin{eqnarray}
 \left\la\lambda_1\left|\mathcal{R}^{\dagger}_M(x_1,k_{1\perp},m_1)
   \right|s_1\right\ra \left\la 00;\frac{1}{2}s_1\left|
   \frac{1}{2}S_z\right\ra\right.=\frac{1}{\sqrt{2(p_1\cdot
   \bar P+m_1M_0)}}\bar{u}(p_1,\lambda_1)\Gamma u(\bar {P},S_z)
\end{eqnarray}
where
\begin{eqnarray}
\Gamma=1, \qquad \qquad \bar {P}=p_1+p_2.
\end{eqnarray}
for the scalar diquark. If the diquark is a vector which is
usually supposed to be the case for the $\Sigma_{c(b)}$ baryon,
the Melosh transform matrix should be modified.

The heavy baryon state is normalized as
 \beq
 \la
 \Lambda(P',S',S'_z)|\Lambda(P,S,S_z)\ra=2(2\pi)^3P^+
  \delta^3(\tilde{P}'-\tilde{P})\delta_{S'S}\delta_{S'_zS_z}.
 \eeq
Thus, the light-front wave function satisfies the constraint
 \beq
 \int\frac{dxd^2k_{\perp}}{2(2\pi^3)}|\phi(x,k_{\perp})|^2=1.
 \eeq

In principle, the wave functions can be obtained by solving the
light-front bound state equations. However, it is too hard to
calculate them based on the first principle, so that instead, we
utilize a phenomenological function, and the Gaussian form would
be most preferable one,
 \beq
 \phi(x,k_{\perp})=N\sqrt{\frac{\partial k_{2z}}{\partial x_2}}
  {\rm exp}\left( \frac{-\vec k^2}{2\beta^2}\right).
 \eeq
with
 \beq
 N=4\left(\frac{\pi}{\beta^2}\right)^{3/4},\qquad
 \frac{\partial k_{2z}}{\partial x_2}=\frac{e_1e_2}{x_1x_2M_0}.
 \eeq
where $\beta$ determines the confinement scale of hadron. The
phenomenological parameters in the light-front quark model are
quark masses and the hadron wave function parameter $\beta$ which
should be prior determined before numerical computations can be
carried out and we will do the job in the later subsections.

\subsection{$\Lambda_b\rightarrow \Lambda_c$ transition form factors}

The form factors for the weak transition $\Lambda_Q\rightarrow
\Lambda_{Q'}$  are defined in the standard way as
\begin{eqnarray}\label{s1}
&& \la \Lambda_{Q'}(P',S',S_z') \mid \bar{Q}'\gamma_{\mu}
 (1-\gamma_{5})Q \mid \Lambda_{Q}(P,S,S_z) \ra  \non \\
 &=& \bar{u}_{\Lambda_{Q'}}(P',S'_z) \left[ \gamma_{\mu} f_{1}(q^{2})
 +i\sigma_{\mu \nu} \frac{ q^{\nu}}{M_{\Lambda_{Q}}}f_{2}(q^{2})
 +\frac{q_{\mu}}{M_{\Lambda_{Q}}} f_{3}(q^{2})
 \right] u_{\Lambda_{Q}}(P,S_z) \nonumber \\
 &&-\bar u_{\Lambda_{Q'}}(P',S'_z)\left[\gamma_{\mu} g_{1}(q^{2})
  +i\sigma_{\mu \nu} \frac{ q^{\nu}}{M_{\Lambda_{Q}}}g_{2}(q^{2})+
  \frac{q_{\mu}}{M_{\Lambda_{Q}}}g_{3}(q^{2})
 \right]\gamma_{5} u_{\Lambda_{Q'}}(P,S_z).
\end{eqnarray}
where  $q \equiv P-P'$, $Q$ and $Q'$ denote $b$ and $c$,
respectively. The above formulation is the most general expression
with only constraint of keeping the Lorentz invariance and parity
conservation for strong interactions. There are six form factors
$f_i,~g_i$ in total for the vector and axial vector current $\bar
c\gamma_{\mu}(1-\gamma_5)b$ and all the information on the strong
interaction is involved in them. Since $S=S'=1/2$, we will be able
to write $\mid\Lambda_{Q}(P,S,S'_z)\ra$ as
$\mid\Lambda_{Q}(P,S_z)\ra$ in the following formulations. A
parametrization is more convenient for the heavy-to-heavy
transitions, which is written in terms of the four-velocities as
 \beq \label{s22}
 &&\la \Lambda_{Q'}(v',S_z') \mid \bar{Q}'\gamma_{\mu}
  (1-\gamma_{5})Q \mid \Lambda_{Q}(v,S_z) \ra  \non \\
  &=& \bar{u}_{\Lambda_{Q'}}(v',S'_z)\left[F_1(\omega)\gamma_{\mu}
  +F_2(\omega)v_{\mu}+F_3(\omega)v_{\mu}^{\prime}\right]
  u_{\Lambda_{Q}}(v,S_z)- \nonumber \\
 &&\bar u_{\Lambda_{Q'}}(v',S'_z)\left[G_1(\omega)\gamma_{\mu}
  +G_2(\omega)v_{\mu}+G_3(\omega)v_{\mu}^{\prime}\right]
 \gamma_{5} u_{\Lambda_{Q}}(v,S_z),
 \eeq
where $v=\frac{P}{M_{\Lambda_Q}},
~v'=\frac{P'}{M_{\Lambda_{Q'}}}$, and $\omega=v\cdot v'$. The
relation between the two methods is
 \beq
 F_1=f_1-\frac{M_{\Lambda_Q}+M_{\Lambda_{Q'}}}{M_{\Lambda_Q}}f_2,
  \qquad F_2=f_2+f_3, \qquad
  F_3=\frac{M_{\Lambda_{Q'}}}{M_{\Lambda_Q}}(f_2-f_3), \non\\
 G_1=g_1+\frac{M_{\Lambda_Q}-M_{\Lambda_{Q'}}}{M_{\Lambda_Q}}g_2,
  \qquad G_2=g_2+g_3, \qquad
  G_3=\frac{M_{\Lambda_{Q'}}}{M_{\Lambda_Q}}(g_2-g_3).
 \eeq

\begin{figure}
\begin{center}
\scalebox{0.8}{\includegraphics{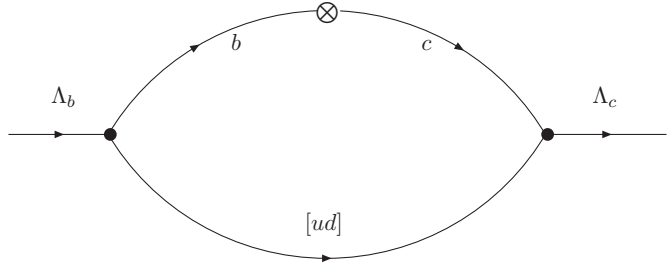}}
\end{center}
\caption{Feynman diagram for $\Lambda_{b}\to\Lambda_{c}$
transitions, where $\bigotimes$ denotes $V-A$ current
vertex.}\label{t1}
\end{figure}

The lowest order Feynman diagram for the $\Lambda_{b}\to\Lambda_{c}$
weak decay is depicted in Fig. \ref{t1}.  In \cite{pentaquark2}, the
light-front quark model for heavy pentaquark with one heavy quark
and two light diquarks is presented. In our case, the heavy baryon
$\Lambda_{b(c)}$ is composed of a heavy quark b(c) and one diquark
[ud]. Thus, most of our formulations are similar to that in
\cite{pentaquark2} except there is only one diquark in our case.

Now, we are ready to calculate the weak transition matrix
elements. Using the the light-front quark model description of
$\mid \Lambda_{Q}(P,S,S_z) \ra$, we obtain
\begin{eqnarray}\label{s2}
&& \la \Lambda_{Q'}(P',S_z') \mid \bar{Q}'
\gamma^{\mu} (1-\gamma_{5}) Q \mid \Lambda_{Q}(P,S_z) \ra  \nonumber \\
 &=& \int\{d^3p_2\}\frac{\phi_{\Lambda_{Q'}}^*(x',k'_{\perp})
  \phi_{\Lambda_Q}(x,k_{\perp})}{2\sqrt{p^+_1p'^+_1(p_1\cdot \bar{P}+m_1M_0)
 (p'_1\cdot \bar{P'}+m'_1M'_0)}}\nonumber \\
  &&\times \bar{u}(\bar{P'},S'_z)\bar{\Gamma}'_{Lm}
  (p_1\!\!\!\!\!\slash'+m'_1)\gamma^{\mu}(1-\gamma_{5})
  (p_1\!\!\!\!\!\slash+m_1)\Gamma u(\bar{P},S_z),
\end{eqnarray}
where
 \beq
 &&\bar{\Gamma}'=\gamma_0\Gamma\gamma_0=\Gamma=1, \non \\
 &&m_1=m_b, \qquad m'_1=m_c, \qquad m_2=m_{[ud]},
 \eeq
and $P$ and $P'$ denote the momenta of initial and final baryons,
$p_1,~p'_1$ are the momenta of $b$ and $c$ quarks, respectively.
Because the diquark is a scalar, one does not need to deal with
spinors which make computations more complex. In this framework,
at each effective vertex, only the three-momentum rather than the
four-momentum is conserved, hence
$\tilde{p}_1-\tilde{p}'_1=\tilde{q}$ and
$\tilde{p}_2=\tilde{p}'_2$. From $\tilde{p}_2=\tilde{p}'_2$, we
have
 \beq
 x'=\frac{P^+}{P^{'+}}x, \qquad \qquad
 k'_{\perp}=k_{\perp}+x_2q_{\perp}.
 \eeq
with $x=x_2$, $x'=x'_2$. Thus, Eq. (\ref{s2}) is rewritten as
\begin{eqnarray}\label{s23}
 &&\la \Lambda_{Q'}(P',S_z') \mid \bar{Q}' \gamma^{\mu}
  (1-\gamma_{5}) Q \mid \Lambda_{Q}(P,S_z) \ra \non\\
  &=& \int\frac{dxd^2k_{\perp}}{2(2\pi)^3}\frac{
  \phi_{\Lambda_{Q'}}(x',k'_{\perp})
  \phi_{\Lambda_Q}(x,k_{\perp})}
  {2\sqrt{x_1x'_1(p_1\cdot \bar{P}+m_1M_0)
  (p'_1\cdot \bar{P'}+m'_1M'_0)}}\nonumber \\
 &&\times \bar{u}(\bar{P'},S'_z)
  (p_1\!\!\!\!\!\slash'+m'_1)\gamma^{\mu}(1-\gamma_{5})
  (p_1\!\!\!\!\!\slash+m_1) u(\bar{P},S_z).
\end{eqnarray}

Following \cite{BD}, the Dirac and Pauli form factors can be derived
from the helicity-conserving and helicity-flip matrix elements of
the plus component of the vector current operators in the
light-front framework. Analogously, the form factors corresponding
to axial vector current are obtained by the authors of
\cite{Schlumpf}. In this work we choose the transverse frame where
$q^+=0,~q_{\perp}\neq0$ which is similar to the treatment in
\cite{pentaquark2}. We then have
 \beq
 f_1(q^2)&=&\frac{\la \Lambda_{Q'}(P',\uparrow)\mid V^+\mid
  \Lambda_{Q}(P,\uparrow)\ra}{2\sqrt{P^+P'^+}}, \non \\
 \frac{f_2(q^2)}{M_{\Lambda_Q}}&=&-\frac{\la \Lambda_{Q'}(P',\uparrow)
  \mid V^+\mid\Lambda_{Q}(P,\downarrow)\ra}{2q_{\perp L}\sqrt{P^+P'^+}},
  \non \\
 g_1(q^2)&=&\frac{\la \Lambda_{Q'}(P',\uparrow)\mid A^+\mid
  \Lambda_{Q}(P,\uparrow)\ra}{2\sqrt{P^+P'^+}},\non\\
 \frac{g_2(q^2)}{M_{\Lambda_Q}}&=&-\frac{\la \Lambda_{Q'}(P',\uparrow)
  \mid A^+\mid\Lambda_{Q}(P,\downarrow)\ra}{2q_{\perp L}\sqrt{P^+P'^+}},
 \eeq
where $q_{\perp L}=q^1_{\perp}-iq^2_{\perp}$.  The above relations
can be written in a compact form as
\begin{eqnarray}\label{s3}
 \la \Lambda_{Q'}(P',S_z') \mid V^+\mid \Lambda_{Q}(P,S_z) \ra
  &=&2\sqrt{P^+P'^+}\left[f_1(q^2)\delta_{S'_zS_z}+
  \frac{f_2(q^2)}{M_{\Lambda_Q}}(\vec{\sigma}\cdot\vec{q}_{\perp}
  \sigma^3)_{S'_zS_z}\right], \nonumber \\
 \la \Lambda_{Q'}(P',S_z') \mid A^+\mid \Lambda_{Q}(P,S_z) \ra
  &=&2\sqrt{P^+P'^+}\left[g_1(q^2)(\sigma^3)_{S'_zS_z}+
  \frac{g_2(q^2)}{M_{\Lambda_Q}}(\vec{\sigma}\cdot
  \vec{q}_{\perp})_{S'_zS_z}\right].
\end{eqnarray}
It is noted that the form factors $f_3(q^2)$ and $g_3(q^2)$ cannot
be extracted in terms of the above method for we have imposed the
condition $q^+=0$. However, they do not contribute to the
semi-leptonic decays $\Lambda_b\to\Lambda_c l\bar\nu_l$ and
vanish in the heavy quark limit.

In order to extract $f_{1,2}(q^2)$ and $g_{1,2}(q^2)$ from Eq.
(\ref{s23}), the following identities are necessary,
\begin{eqnarray}\label{s4}
 \frac{1}{2}\sum_{S_z,S'_z}u(\bar{P},S_z)\delta_{S'_zS_z}\bar{u}
  (\bar{P}',S'_z)&=&\frac{1}{4\sqrt{P^+P'^+}}
  (\bar{P}\!\!\!\!\slash+M_0)\gamma^+(\bar{P'}\!\!\!\!\!\slash+M'_0),
  \nonumber \\
 \frac{1}{2}\sum_{S_z,S'_z}u(\bar{P},S_z)
  (\sigma^3\sigma^i_{\perp})_{S'_zS_z}\bar{u}
  (\bar{P}',S'_z)&=&-\frac{1}{4\sqrt{P^+P'^+}}
  (\bar{P}\!\!\!\!\slash+M_0)\sigma^{i+}(\bar{P'}\!\!\!\!\!\slash+
   M'_0), \nonumber\\
 \frac{1}{2}\sum_{S_z,S'_z}u(\bar{P},S_z)(\sigma^3)_{S'_zS_z}\bar{u}
  (\bar{P}',S'_z)&=&\frac{1}{4\sqrt{P^+P'^+}}
  (\bar{P}\!\!\!\!\slash+M_0)\gamma^{+}\gamma_5(\bar{P'}
   \!\!\!\!\!\slash+M'_0), \nonumber\\
 \frac{1}{2}\sum_{S_z,S'_z}u(\bar{P},S_z)(\sigma^i_{\perp})_{S'_zS_z}
  \bar{u}(\bar{P}',S'_z)&=&-\frac{1}{4\sqrt{P^+P'^+}}
 (\bar{P}\!\!\!\!\slash+M_0)\sigma^{i+}\gamma_5(\bar{P'}
  \!\!\!\!\!\slash+M'_0).
\end{eqnarray}
It should be noted that $u(\bar P,S_z)$ is not equal to
$u(P,S_z)$, but the relation $\gamma^+u(\bar
P,S_z)=\gamma^+u(P,S_z)$ is satisfied.

From Eqs. (\ref{s23}), (\ref{s3}) and (\ref{s4}), the transition
form factors are obtained,
\begin{eqnarray}\label{s6}
 f_1(q^2)&=&\frac{1}{8P^+P'^+}\int{\frac{dxd^2k_{\perp}}{2(2\pi)^3}}
  \frac{\phi_{\Lambda_{Q'}}(x',k'_{\perp})
  \phi_{\Lambda_Q}(x,k_{\perp})}{2\sqrt{x_1x'_1
  (p_1\cdot\bar{P}+m_1M_0)(p'_1\cdot \bar{P'}+m'_1M'_0)}} \non\\
 &&\times {\rm Tr}\left[(\bar{P}\!\!\!\!\slash+M_0)\gamma^+
  (\bar{P'}\!\!\!\!\!\slash+M'_0)(p_1\!\!\!\!\!\slash'+m'_1)
  \gamma^{+}(p_1\!\!\!\!\!\slash+m_1)\right],\non \\
 g_1(q^2)&=&\frac{1}{8P^+P'^+}\int{\frac{dxd^2k_{\perp}}{2(2\pi)^3}}
  \frac{\phi_{\Lambda_{Q'}}(x',k'_{\perp})
  \phi_{\Lambda_Q}(x,k_{\perp})}{2\sqrt{x_1x'_1
  (p_1\cdot\bar{P}+m_1M_0)(p'_1\cdot \bar{P'}+m'_1M'_0)}} \non\\
 &&\times {\rm Tr}\left[(\bar{P}\!\!\!\!\slash+M_0)\gamma^+\gamma_5
  (\bar{P'}\!\!\!\!\!\slash+M'_0)
  (p_1\!\!\!\!\!\slash'+m'_1)\gamma^{+}\gamma_5
  (p_1\!\!\!\!\!\slash+m_1)\right],\nonumber \\
 \frac{f_2(q^2)}{M_{\Lambda_Q}}&=&-\frac{1}{8P^+P'^+q^i_{\perp}}
  \int{\frac{dxd^2k_{\perp}}{2(2\pi)^3}}
  \frac{\phi_{\Lambda_{Q'}}(x',k'_{\perp})
  \phi_{\Lambda_Q}(x,k_{\perp})}{2\sqrt{x_1x'_1
  (p_1\cdot\bar{P}+m_1M_0)(p'_1\cdot \bar{P'}+m'_1M'_0)}} \non \\
 &&\times{\rm Tr}\left[(\bar{P}\!\!\!\!\slash+M_0)\sigma^{i+}
  (\bar{P'}\!\!\!\!\!\slash+M'_0)
  (p_1\!\!\!\!\!\slash'+m'_1)\gamma^{+}
  (p_1\!\!\!\!\!\slash+m_1)\right],\non \\
 \frac{g_2(q^2)}{M_{\Lambda_Q}}&=&\frac{1}{8P^+P'^+q^i_{\perp}}
  \int{\frac{dxd^2k_{\perp}}{2(2\pi)^3}}
  \frac{\phi_{\Lambda_{Q'}}(x',k'_{\perp})
  \phi_{\Lambda_Q}(x,k_{\perp})}{2\sqrt{x_1x'_1
  (p_1\cdot\bar{P}+m_1M_0)(p'_1\cdot \bar{P'}+m'_1M'_0)}} \non \\
 &&\times{\rm Tr}\left[(\bar{P}\!\!\!\!\slash+M_0)\sigma^{i+}\gamma_5
  (\bar{P'}\!\!\!\!\!\slash+M'_0)
  (p_1\!\!\!\!\!\slash'+m'_1)\gamma^{+}\gamma_5
  (p_1\!\!\!\!\!\slash+m_1)\right],
\end{eqnarray}
with $i=1,2$. The traces can be worked out straightforwardly
\begin{eqnarray}
 &&\frac{1}{8P^+P'^+}{\rm Tr}\left[(\bar{P}\!\!\!\!\slash+M_0)
   \gamma^+(\bar{P'}\!\!\!\!\!\slash+M'_0)
   (p_1\!\!\!\!\!\slash'+m'_1)\gamma^{+}
   (p_1\!\!\!\!\!\slash+m_1)\right]\nonumber \\
 &&~~~~~=-(p_1-x_1\bar{P})\cdot(p'_1-x'_1\bar{P}')+
   (x_1M_0+m_1)(x'_1M'_0+m'_1),\nonumber \\
 &&\frac{1}{8P^+P'^+}{\rm Tr}\left[(\bar{P}\!\!\!\!\slash+M_0)
   \gamma^+\gamma_5(\bar{P'}\!\!\!\!\!\slash+M'_0)
   (p_1\!\!\!\!\!\slash'+m'_1)\gamma^{+}\gamma_5
   (p_1\!\!\!\!\!\slash+m_1)\right]\nonumber \\
 &&~~~~~=(p_1-x_1\bar{P})\cdot(p'_1-x'_1\bar{P}')+
   (x_1M_0+m_1)(x'_1M'_0+m'_1),\nonumber
 \eeq
and
 \beq\label{s7}
 &&\frac{1}{8P^+P'^+}{\rm Tr}\left[(\bar{P}\!\!\!\!\slash+M_0)\sigma^{i+}
   (\bar{P'}\!\!\!\!\!\slash+M'_0)(p_1\!\!\!\!\!\slash'+m'_1)\gamma^{+}
   (p_1\!\!\!\!\!\slash+m_1)\right]\non\\
 &&~~~~~=(m'_1+x'_1\bar{M}'_0)(p^i_{1\perp}- x_1\bar{P}^i_{\perp})-
   (m_1+x_1M_0)(p^{\prime i}_{1\perp}- x'_1\bar{P}'^i_{\perp}),\nonumber\\
 &&\frac{1}{8P^+P'^+}{\rm Tr}\left[(\bar{P}\!\!\!\!\slash+M_0)\sigma^{i+}\gamma_5
   (\bar{P'}\!\!\!\!\!\slash+M'_0)(p_1\!\!\!\!\!\slash'+m'_1)\gamma^{+}\gamma_5
   (p_1\!\!\!\!\!\slash+m_1)\right]\nonumber\\
 &&~~~~~=(m'_1+x'_1\bar{M}'_0)(p^i_{1\perp}-x_1\bar{P}^i_{\perp})+(m_1+x_1M_0)
   (p^{\prime i}_{1\perp}-x'_1\bar{P}'^i_{\perp}).
\end{eqnarray}
Using $\bar P^{(\prime)}=p_1^{(\prime)}+p_2^{(\prime)}$ and other
momentum relations, the products of momenta in Eqs. (\ref{s6}) and
(\ref{s7}) are given in terms of the internal variables as
\begin{eqnarray}\label{s8}
 &&p_1\cdot\bar{P}=e_1M_0=
   \frac{m^2_1+x^2_1M^2_0+k_{1\perp}^2}{2x_1}, \non\\
 &&p'_1\cdot\bar{P}'=e'_1M'_0=
   \frac{m'^2_1+x'^2_1M'^2_0+{k'}_{1\perp}^{2}}{2x'_1},\non\\
 &&p_{1\perp}^{(\prime)i}-x_1\bar{P}_{\perp}^{(\prime)i}=
   k_{1\perp}^{(\prime)i}, \non\\
 &&(p_1-x_1\bar{P})\cdot(p'_1-x'_1\bar{P}')=
   -k_{1\perp}\cdot k'_{1\perp}.
\end{eqnarray}

At last, we obtain the final expressions for the
$\Lambda_Q\to\Lambda_{Q'}$ weak transition form factors
\begin{eqnarray}\label{s9}
 f_1(q^2)&=&\int{\frac{dxd^2k_{\perp}}{2(2\pi)^3}}
   \frac{\phi_{\Lambda_{Q'}}(x',k'_{\perp})
  \phi_{\Lambda_Q}(x,k_{\perp})\left[k_{2\perp}
   \cdot k'_{2\perp}+\left(x_1M_0+m_1\right)
   \left(x'_1M'_0+m'_1\right)\right]}
   {\sqrt{\left[\left(m_1+x_1M_0\right)^2+k_{2\perp}^2\right]
   \left[\left(m'_1+x_1M'_0\right)^2+k_{2\perp}^{'2}\right]}}, \non \\
 g_1(q^2)&=&\int{\frac{dxd^2k_{\perp}}{2(2\pi)^3}}
   \frac{\phi_{\Lambda_{Q'}}(x',k'_{\perp})
  \phi_{\Lambda_Q}(x,k_{\perp})[-k_{2\perp}
   \cdot k'_{2\perp}+(x_1M_0+m_1)(x'_1M'_0+m'_1)]}
   {\sqrt{\left[\left(m_1+x_1M_0\right)^2+k_{2\perp}^2\right]
   \left[\left(m'_1+x_1M'_0\right)^2+k_{2\perp}^{'2}\right]}}, \non \\
 \frac{f_2(q^2)}{M_{\Lambda_Q}}&=&\frac{1}{q^i_{\perp}}
   \int{\frac{dxd^2k_{\perp}}{2(2\pi)^3}}
   \frac{\phi_{\Lambda_{Q'}}(x',k'_{\perp})
   \phi_{\Lambda_Q}(x,k_{\perp})
   [(m_1+x_1M_0)k_{1\perp}^{\prime i}-(m'_1+x'_1M'_0)k_{1\perp}^i]}
   {\sqrt{\left[\left(m_1+x_1M_0\right)^2+k_{2\perp}^2\right]
   \left[\left(m'_1+x_1M'_0\right)^2+k_{2\perp}^{'2}\right]}}, \non \\
 \frac{g_2(q^2)}{M_{\Lambda_Q}}&=&\frac{1}{q^i_{\perp}}
   \int{\frac{dxd^2k_{\perp}}{2(2\pi)^3}}
   \frac{\phi_{\Lambda_{Q'}}(x',k'_{\perp})
   \phi_{\Lambda_Q}(x,k_{\perp})
   [(m_1+x_1M_0)k_{1\perp}^{\prime i}+ (m'_1+x'_1M'_0)
   {k}_{1\perp}^i]}{\sqrt{\left[\left(m_1+x_1M_0\right)^2+
   k_{2\perp}^2\right]\left[\left(m'_1+x_1M'_0\right)^2+
   k_{2\perp}^{'2}\right]}}.\non\\
\end{eqnarray}

\subsection{The Form factors in the heavy quark limit}

It is well known that there is a non-trivial symmetry in QCD: the
heavy quark symmetry (HQS) in the infinite quark mass limit
\cite{HQS}. Since the masses of heavy quarks $b$ and $c$ are much
larger than the strong interaction scale $\Lambda_{\rm QCD}$, the
spin of the heavy quark decouples from light quark and gluon
degrees of freedoms, and an extra symmetry $SU_f(2)\otimes
SU_s(2)$ is expected. This flavor and spin symmetry provides
several model-independent relations for the heavy-to-heavy
baryonic form factors: the six form factors $f_i,~g_i$ (i=1,2,3)
are related to a unique universal Isgur-Wise function
$\zeta(v\cdot v')$. In the heavy quark limit, the heavy quark $Q$
is described by a two-component spinor $Q_v={\rm e}^{im_Qv\cdot
x}\frac{(1+\vslash)}{2}Q$ where $v$ is the velocity of the heavy
baryon. The current $\bar Q'\gamma_{\mu}(1-\gamma_5)Q$ in the full
theory is matched onto the current $\bar{Q}_{v'}' \gamma_{\mu}
(1-\gamma_{5}) Q_v$ in the heavy quark effective theory (HQET).
The baryon bound state and Dirac spinor field are replaced by
 \beq
 \mid\Lambda_{Q}(P,S_z)\ra &\to& \sqrt{M_{\Lambda_Q}}
 \mid\Lambda_{Q}(v,S_z)\ra,
  \non \\
 u(\bar{P},S_z) &\to& \sqrt{m_Q}u(v,S_z).
 \eeq
The Isgur-Wise function which appears in the transition amplitude
$\Lambda_Q\to\Lambda_{Q'}$ is defined as \cite{IW}
 \begin{eqnarray}\label{s10}
  \la \Lambda_{Q'}(v',S_z')\mid\bar{Q}_{v'}' \gamma^{\mu}
  (1-\gamma_{5}) Q_v \mid \Lambda_{Q}(v,S_z)\ra =
  \zeta(\omega)\bar{u}(v',S'_z)\gamma_{\mu}(1-\gamma_{5})u(v,S_z),
 \end{eqnarray}
where $\omega\equiv v\cdot v'$. The heavy flavor symmetry implies
that the Isgur-Wise function is normalized to be 1 at the
zero-recoil point, $\zeta(1)=1$. The physical form factors are
obtained as
 \begin{eqnarray}\label{s11}
  f_1(q^2)=g_1(q^2)=\zeta(\omega),\qquad \qquad
  f_2=f_3=g_2=g_3=0,
 \end{eqnarray}
where $q^2=M_{\Lambda_Q}^2+M_{\Lambda_{Q'}}^2-
2M_{\Lambda_Q}M_{\Lambda_{Q'}}\omega$.

Since the momentum of $\Lambda_Q$  is dominated by the momentum of
the heavy quark $Q$, the momentum of the light spectator diquark
$x$ is of order $\lqcd/m_Q$. The variable $X\equiv xm_b$ is of
order of $\lqcd$. In analog to the situation for heavy mesons, the
wave function of $\Lambda_Q$ should have a scaling behavior in the
heavy quark limit \cite{CZL}
 \beq
  \phi_{\Lambda_Q}(x,k_{\bot})\to \sqrt{\frac{m_Q}{X}}\Phi(X,k_{\bot}),
 \eeq
where the factor $\sqrt{m_Q}$ is deliberately factorized out and
the rest  of $\phi_{\Lambda_Q}(x,k_{\bot})$  is independent of
$m_Q$,  $\Phi(X,k_{\bot})$ is normalized as
 \beq \label{s24}
 \int_0^{\infty}\frac{dX}{X}\int\frac{d^2k_{\bot}}{2(2\pi)^3}
  |\Phi(X,k_{\bot})|^2=1.
 \eeq
For the on-shell diquark momentum $p_2$, we have
$p^-_2=\frac{(p^2_{2\perp}+m^2_2)}{p^+_2}$, and
 \beq
 v\cdot p_2=\frac{m_2^2+k_{\perp}^2+X^2}{2X}.
 \eeq
Hence the wave function $\Phi(X,k_{\bot})$ which only depends on
the velocity of the baryon, is the same for $\Lambda_Q$ and
$\Lambda_{Q'}$.

We now consider the transition form factors obtained in the
previous section under the heavy quark limit. For the initial
baryon, we have
\begin{eqnarray}\label{s12}
 &&M_{\Lambda_Q}\to m_Q, \qquad ~~ M_0\to m_Q,\non\\
 &&e_1\to m_Q, \qquad\qquad e_2\to v\cdot p_2,\non\\
 &&\vec k^2\to (v\cdot p_2)^2-m_2^2,\non\\
 &&p_1\!\!\!\!\!\slash+m_1\to m_Q(v\!\!\!\slash+1)\non \\
 &&\frac{e_1e_2}{x_1x_2M_0}\to\frac{m_Q}{X}(v\cdot p_2),
\end{eqnarray}
and
 \beq
 \Phi(X,k_{\bot})=4\sqrt{v\cdot p_2}\left(\frac{\pi}{\beta^2_\infty}
  \right)^{\frac{3}{4}}{\rm exp}\left(-\frac{(v\cdot p_2)^2-m^2_2}
  {2\beta^2_\infty}\right).
 \eeq
The subscript in $\beta_{\infty}$ represents the case of the heavy
quark limit. Similar expressions can be obtained for the final
baryon where a prime sign ``$'$" would be attached to each
variable.

The calculation of the Isgur-Wise function in the heavy quark limit
becomes much simpler than that for $f_i$ and $g_i$ because they can
be evaluated directly in the time-like region by choosing a
reference frame where $q_\perp=0$ \cite{pentaquark2}. The  matrix
element $\Lambda_Q\to\Lambda_{Q'}$ is
\begin{eqnarray}\label{sw14}
 &&\la\Lambda_{Q'}(v',S_z') \mid \bar{Q}_{v'}'
   \gamma^{\mu}(1-\gamma_{5})Q_v\mid\Lambda_{Q}(v,S_z) \ra  \non\\
 &=& \int{\frac{dX}{X}\frac{d^2k_{\perp}}{2(2\pi)^3}}
 {\Phi(X,k_{\perp})\Phi(X',k_{\perp}^{\prime})}
 \bar{u}(v',S'_z)\gamma^\mu(1-\gamma_5)u(v,S_z),
\end{eqnarray}
where $z\equiv X'/X$. By comparing the above equation with Eq.
(\ref{s10}), we get
\begin{eqnarray}\label{s15}
 \zeta(\omega)= \int{\frac{dX}{X}\frac{d^2k_{\perp}}{2(2\pi)^3}}
 {\Phi(X,k_{\perp})\Phi(X',k_{\perp}^{\prime})}.
\end{eqnarray}
The obtained Isgur-Wise function $\zeta(\omega)$ is an overlapping
integration of the initial and final wave functions and no spin
information is left. The variable $z$ is related to $\omega$ via
\begin{eqnarray}\label{s16}
 z\rightarrow z_{\pm}=\omega\pm\sqrt{\omega^2-1},
 \qquad\qquad z_+=\frac{1}{z_-}.
\end{eqnarray}
where $+(-)$ denotes the final baryon recoiling direction. There
is a symmetry between $z_+$ and $z_-$. $\zeta(\omega)$ does not
change when we replace $z_+$ by $z_-$, or vice versa. Eq.
(\ref{s15}) shows explicitly that $\zeta(\omega)$ depends only on
the velocities of the initial and final baryons and independent of
the heavy quark masses.

The Isgur-Wise function can also be obtained from Eq. (\ref{s2})
by taking the heavy quark limit. It is not difficult to verify
that $f_1(q^2)=g_1(q^2)=\zeta(\omega)$, $f_2=g_2=0$ in leading
order of $\lqcd/m_Q$. The normalization of Isgur-Wise function at
the zero-recoil point is guaranteed by our normalization condition
for wave functions Eq. (\ref{s24}). Its consistency with forms in
the heavy quark limit implies the correctness of the light-front
approach.

\section{Semi-leptonic and Non-leptonic decays of transition
$\Lambda_b \to \Lambda_c$ }

The polarization effects in exclusive processes, such as $B\to
\phi K^*$ offers non-trivial information about strong interaction,
which is important to test different theoretical approaches. The
decays of $\Lambda_b\to\Lambda_c$ indeed contain complex spin
structures. In this section, we obtain formulations for the rates
of semi-leptonic and non-leptonic processes. In this work, we
concern only the exclusive decay modes.

\subsection{Semi-leptonic decays of
 $\Lambda_b \to \Lambda_c l\bar\nu_l$ }

The transition amplitude of $\Lambda_b\to\Lambda_c$ contains
several independent helicity components. The helicity amplitudes
induced by the weak vector and axial vector currents are described
by $H^{V,A}_{\lambda',\lambda_W}$ where $\lambda'$ and $\lambda_W$
denote the helicities of the final baryon and the virtual
$W$-boson, respectively. According to the definitions of the form
factors for $\Lambda_b \to \Lambda_c$ given in Eq. (\ref{s22}),
the helicity amplitudes are related to these form factors through
the following expressions \cite{KKP}
 \beq
 H^V_{\frac{1}{2},0}&=&\frac{\sqrt{Q_-}}{\sqrt{q^2}}\left(
  \left(\Mb+\Mc\right)f_1-\frac{q^2}{\Mb}f_2\right),\non\\
 H^V_{\frac{1}{2},1}&=&\sqrt{2Q_-}\left(-f_1+
  \frac{\Mb+\Mc}{\Mb}f_2\right),\non\\
 H^A_{\frac{1}{2},0}&=&\frac{\sqrt{Q_+}}{\sqrt{q^2}}\left(
  \left(\Mb-\Mc\right)g_1+\frac{q^2}{\Mb}g_2\right),\non\\
 H^A_{\frac{1}{2},1}&=&\sqrt{2Q_+}\left(-g_1-
  \frac{\Mb-\Mc}{\Mb}g_2\right).
 \eeq
where $Q_{\pm}=2(P\cdot P'\pm \Mb\Mc)=2\Mb\Mc(\omega\pm 1)$. The
amplitudes for the negative helicities are obtained in terms of
the relation
 \beq
 H^{V,A}_{-\lambda'-\lambda_W}=\pm H^{V,A}_{\lambda',\lambda_W},
  \eeq
where the upper (lower) sign corresponds to V(A).

Because of the $V-A$ structure of the weak current, the helicity
amplitudes are obtained as
 \beq
 H_{\lambda',\lambda_W}=H^V_{\lambda',\lambda_W}-
  H^A_{\lambda',\lambda_W}.
 \eeq
The helicities of the $W$-boson $\lambda_W$ can be either $0$ or
$1$, which correspond to the longitudinal and transverse
polarizations, respectively. Following the definitions in
literature, we decompose the decay width into a sum of the
longitudinal and transverse parts according to the helicity states
of the virtual W-boson. The differential decay rate of $\Lambda_b
\to \Lambda_c l\bar\nu_l$ is
 \beq
 \frac{d\Gamma}{d\omega}=\frac{d\Gamma_L}{d\omega}+
 \frac{d\Gamma_T}{d\omega},
 \eeq
and the longitudinally (L) and transversely (T) polarized rates are
respectively\cite{KKP}
 \beq
 \frac{d\Gamma_L}{d\omega}&=&\frac{G_F^2|V_{cb}|^2}{(2\pi)^3}~
  \frac{q^2~p_c~\Mc}{12\Mb}\left[
  |H_{\frac{1}{2},0}|^2+|H_{-\frac{1}{2},0}|^2\right],\non\\
 \frac{d\Gamma_T}{d\omega}&=&\frac{G_F^2|V_{cb}|^2}{(2\pi)^3}~
  \frac{q^2~p_c~\Mc}{12\Mb}\left[
  |H_{\frac{1}{2},1}|^2+|H_{-\frac{1}{2},-1}|^2\right].
 \eeq
where $p_c=\Mc\sqrt{\omega^2-1}$ is the momentum of $\Lambda_c$ in
the reset frame of $\Lambda_b$. Integrating over the solid angle,
we obtain the decay rate
 \beq
 \Gamma=\int_1^{\omega_{\rm max}}d\omega\frac{d\Gamma}{d\omega},
 \eeq
where the upper bound of the integration $\omega_{\rm
max}=\frac{1}{2}\left(\frac{M_{\Lambda_{b}}}
{M_{\Lambda_{c}}}+\frac{M_{\Lambda_{c}}}{M_{\Lambda_{b}}}\right)$
is the maximal recoil. In order to compare our results with those
in the literatures, we used the variable $\omega$ in the
expression for the differential decay rate. In the heavy quark
limit, the decay rate of $\Lambda_b \to \Lambda_c l\bar{\nu}_l$ is
simplified into
\begin{eqnarray}
 \frac{d\Gamma}{d\omega}=\frac{G_F^{2}|V_{cb}|^2}
  {24\pi^3}M_{\Lambda_{b}}^5 \sqrt{\omega^2-1}~r^3 ( 6\omega
  r^2-8\omega^2r-4r+6\omega)\zeta(\omega)^2,
\end{eqnarray}
with $r=\frac{M_{\Lambda_c}}{M_{\Lambda_b}}$.

The polarization of the cascade decay $\Lambda_b\to\Lambda_c(\to
p\pi)+W(\to l\nu)$ is expressed by various asymmetry
parameters\cite{EFG,KKP}. Among them, the integrated longitudinal
and transverse asymmetries are defined by
 \beq
 a_L&=&\frac{\int_1^{\omega_{\rm max}} d\omega ~q^2~ p_c
     \left[ |H_{\frac{1}{2},0}|^2-|H_{-\frac{1}{2},0}|^2\right]}
     {\int_1^{\omega_{\rm max}} d\omega ~q^2~ p_c
     \left[|H_{\frac{1}{2},0}|^2+|H_{-\frac{1}{2},0}|^2\right]},
     \non\\
 a_T&=&\frac{\int_1^{\omega_{\rm max}} d\omega ~q^2~ p_c
     \left[ |H_{\frac{1}{2},1}|^2-|H_{-\frac{1}{2},-1}|^2\right]}
     {\int_1^{\omega_{\rm max}} d\omega ~q^2~ p_c
     \left[|H_{\frac{1}{2},1}|^2+|H_{-\frac{1}{2},-1}|^2\right]}.
 \eeq
The ratio of the longitudinal to transverse decay rates $R$ is
defined by
 \beq
 R=\frac{\Gamma_L}{\Gamma_T}=\frac{\int_1^{\omega_{\rm
     max}}d\omega~q^2~p_c\left[ |H_{\frac{1}{2},0}|^2+|H_{-\frac{1}{2},0}|^2
     \right]}{\int_1^{\omega_{\rm max}}d\omega~q^2~p_c
     \left[ |H_{\frac{1}{2},1}|^2+|H_{-\frac{1}{2},-1}|^2\right]},
 \eeq
and the  longitudinal $\Lambda_c$ polarization asymmetry $P_L$ is
given as
 \beq
 P_L&=&\frac{\int_1^{\omega_{\rm max}} d\omega ~q^2~ p_c
     \left[ |H_{\frac{1}{2},0}|^2-|H_{-\frac{1}{2},0}|^2+
     |H_{\frac{1}{2},1}|^2-|H_{-\frac{1}{2},-1}|^2\right]}
     {\int_1^{\omega_{\rm max}} d\omega ~q^2~ p_c
     \left[|H_{\frac{1}{2},0}|^2+|H_{-\frac{1}{2},0}|^2+
     |H_{\frac{1}{2},1}|^2+|H_{-\frac{1}{2},-1}|^2\right]}
  \non\\
  &=&\frac{a_T+R\,a_L}{1+R}.
 \eeq

\subsection{Non-leptonic decay  of $\Lambda_b \to \Lambda_c M$}

Several exclusive non-leptonic decays of $\Lambda_b \to \Lambda_c
+M$ where $M$ is a meson, have been measured in recent experiments
\cite{PDG06}. From the theoretical aspects, the non-leptonic
decays are much more complicated than the semi-leptonic ones
because of the strong interaction. Generally, the present
theoretical framework is based on the factorization assumption,
where  the hadronic matrix element is factorized into a product of
two matrix elements of single currents. One can be written as a
decay constant while the other is expressed in terms of a few form
factors according to the lorentz structure of the current. For the
weak decays of mesons, such factorization approach is verified to
work very well for the color-allowed processes and the
non-factorizable contributions are negligible. We have reason to
believe that this would be valid for the baryon case, especially
as the diquark picture is employed. The decays $\Lambda_b^0 \to
\Lambda_c^+ M^-$ belong to this type. Thus, the study on these
modes could be not only a test for the factorization hypothesis,
but also a check of the consistency of the obtained form factors
in the heavy bottomed baryon system.

For the non-leptonic decays $\Lambda_b^0 \to \Lambda_c^+ M^-$, the
effective interaction at the quark level is $b\to c\bar{q_1}q_2$.
The relevant Hamiltonian is
 \beq
 &&{\cal H}_W=\frac{G_F}{\sqrt 2}V_{cb}V_{q_1q_2}^*(c_1O_1+c_2O_2),
  \non\\
 &&O_1=(\bar c b)_{V-A} (\bar q_2q_1)_{V-A},\qquad
 O_2=(\bar q_2b)_{V-A} (\bar c q_1)_{V-A}.
 \eeq
where $c_i$ denotes the short-distance Wilson coefficient,
$V_{cb}(V_{q_1q_2})$ is the CKM matrix elements, $q_1$ stands for
$u$ or $c$ and $q_2$ for $d$ or $s$ in the context. Then one needs
to evaluate the hadronic matrix elements
 \beq
 \la \Lambda_c M | {\cal H}_W | \Lambda_b\ra=
 \frac{G_F}{\sqrt 2}V_{cb}V_{q_1q_2}^*\sum_{i=1,2}c_i~
 \la \Lambda_c M | O_i | \Lambda_b\ra.
 \eeq
Under the factorization approximation, the hadronic matrix element
is reduced to
 \beq
 \la\Lambda_c M | O_i | \Lambda_b\ra
 =\la\Lambda_c | J_\mu |\Lambda_b\ra
  \la M | J^{\prime\mu} | 0\ra.
 \eeq
where $J(J')$ is the $V-A$ weak current. The first factor
$\la\Lambda_c | J_\mu |\Lambda_b\ra$ is parameterized by six form
factors as done in Eq. (\ref{s1}). The second factor defines the
decay constants as follows
 \beq
 \la P(P)|A_{\mu}|0\ra&=&f_PP_{\mu}, \non\\
 \la S(P)|V_{\mu}|0\ra&=&f_SP_{\mu}, \non\\
 \la V(P,\epsilon)|V_{\mu}|0\ra&=&f_VM_V\epsilon^*_{\mu}, \non\\
 \la A(P,\epsilon)|A_{\mu}|0\ra&=&f_VM_A\epsilon^*_{\mu},
 \eeq
where $P(V)$ denotes a pseudoscalar (vector) meson, and $S(A)$
denotes a scalar (axial-vector) meson. In the definitions, we omit
a factor $(-i)$ for the  pseudoscalar meson decay constant.

In general, the transition amplitude of $\Lambda_b\to\Lambda_c M$
can be written as
 \beq
 {\cal M}(\Lambda_b\to\Lambda_c P)&=&\bar
  u_{\Lambda_c}(A+B\gamma_5)u_{\Lambda_b}, \non \\
 {\cal M}(\Lambda_b\to\Lambda_c V)&=&\bar
  u_{\Lambda_c}\epsilon^{*\mu}\left[A_1\gamma_{\mu}\gamma_5+
   A_2(p_{\Lambda_c})_{\mu}\gamma_5+B_1\gamma_{\mu}+
   B_2(p_{\Lambda_c})_{\mu}\right]u_{\Lambda_b},
 \eeq
where $\epsilon^{\mu}$ is the polarization vector of the final
vector or axial-vector mesons. Including the effective Wilson
coefficient $a_1=c_1+c_2/N_c$, the decay amplitudes in the
factorization approximation are \cite{KK,Cheng}
 \beq
 A&=&\lambda f_P(\Mb-\Mc)f_1(M^2), \non \\
 B&=&\lambda f_P(\Mb+\Mc)g_1(M^2), \non\\
 A_1&=&-\lambda f_VM\left[g_1(M^2)+g_2(M^2)\frac{\Mb-\Mc}{\Mb}\right],
 \non\\
 A_2&=&-2\lambda f_VM\frac{g_2(M^2)}{\Mb},\non\\
 B_1&=&\lambda f_VM\left[f_1(M^2)-f_2(M^2)\frac{\Mb+\Mc}{\Mb}\right],
 \non\\
 B_2&=&2\lambda f_VM\frac{f_2(M^2)}{\Mb},
 \eeq
where $\lambda=\frac{G_F}{\sqrt 2}V_{cb}V_{q_1q_2}^*a_1$ and $M$ is
the meson mass. Replacing  $P$, $V$ by $S$ and $A$ in the above
expressions, one can easily obtain similar expressions for scalar
and axial-vector mesons .

The decay rates of $\Lambda_b\rightarrow\Lambda_cP(S)$ and up-down
asymmetries are \cite{Cheng}
 \begin{eqnarray}
 \Gamma&=&\frac{p_c}{8\pi}\left[\frac{(\Mb+\Mc)^2-M^2}{\Mb^2}|A|^2+
  \frac{(\Mb-\Mc)^2-M^2}{\Mb^2}|B|^2\right], \non\\
 \alpha&=&-\frac{2\kappa{\rm Re}(A^*B)}{|A|^2+\kappa^2|B|^2},
 \end{eqnarray}
where $p_c$ is the $\Lambda_c$ momentum in the rest frame of
$\Lambda_b$ and $\kappa=\frac{p_c}{E_{\Lambda_c}+\Mc}$. For
$\Lambda_b\rightarrow\Lambda_c V(A)$ decays, the decay rates and
up-down asymmetries are
 \beq
 \Gamma&=&\frac{p_c (E_{\Lambda_c}+\Mc)}{8\pi\Mb}\left[
  2\left(|S|^2+|P_2|^2\right)+\frac{E^2}{M^2}\left(
  |S+D|^2+|P_1|^2\right)\right], \non\\
 \alpha&=&\frac{4M^2{\rm Re}(S^*P_2)+2E^2{\rm Re}(S+D)^*P_1}
  {2M^2\left(|S|^2+|P_2|^2 \right)+E^2\left(|S+D|^2+|P_1|^2
  \right) },
 \eeq
where $E$ is energy of the vector (axial vector) meson, and
 \begin{eqnarray}
  S&=&-A_1, \non\\
  P_1&=&-\frac{p_c}{E}\left(\frac{\Mb+\Mc}
  {E_{\Lambda_c}+\Mc}B_1+B_2\right), \non \\
  P_2&=&\frac{p_c}{E_{\Lambda_c}+\Mc}B_1,\non\\
  D&=&-\frac{p^2_c}{E(E_{\Lambda_c}+\Mc)}(A_1- A_2).
 \end{eqnarray}

\section{Numerical Results}

In this section, we will present our numerical results of  the
form factors for the transition $\Lambda_b\to\Lambda_c$. Then use
them to predict the rates of the exclusive semi-leptonic
$\Lambda_b\rightarrow \Lambda_c l\bar{\nu}_l$, and two-body
non-leptonic processes, such as $\Lambda_b\to \Lambda_c M^-$ where
$M=\pi,~ K,~ \rho,~ K^*,~ a_1$.

At first, we provide our input parameters in the light-front quark
model. The baryon masses $\Mb=5.624$ GeV, $\Mc=2.285$ GeV are taken
from \cite{PDG06}. The quark masses and the hadron wave function
parameter $\beta$ need to be specified. For the heavy quark masses,
we take $m_b$ and $m_c$ from \cite{CCH2}. Following
\cite{pentaquark2}, the mass of a [ud] diquark is assumed to be
close to the constitute strange quark mass. In the literature, the
mass of the constitutent light scalar diquark $m_{[ud]}$ is rather
arbitrary, for example, it is set as: 400 MeV \cite{pentaquark2},
500 MeV \cite{MQS}, 710 MeV \cite{EFG} and 650$\sim$800 MeV
\cite{Guo}. A recent result from lattice calculation gives the
scalar diquark mass varies from 1190 to 696 MeV when a hopping
parameter $\kappa$ changes from 0.140 to 0.148 \cite{HKLW}. To
reduce error and model-dependence, we use the value of $BR(\Lambda_b
\to \Lambda_c l\bar{\nu}_l)=5.0^{+1.1}_{-0.8}({\rm
stat})^{+1.6}_{-1.2}({\rm syst})\%$ measured by the DELPHI
collaboration \cite{DELPHI} to fix parameters. The present data
favors diquark mass as $m_{[ud]}=500$ MeV. All the input parameters
are collected in Table \ref{Tab:t1}.

\begin{table}
\caption{Quark mass and the parameter $\beta$ (in units of
 GeV).}\label{Tab:t1}
\begin{ruledtabular}
\begin{tabular}{ccccc}
  $m_c$  & $m_b$  &$m_{[ud]}$ & $\beta_{c[ud]}$ & $\beta_{b[ud]}$ \\
  $1.3$  & $4.4$  & 0.50     & $0.35$         & 0.40
\end{tabular}
\end{ruledtabular}
\end{table}

\subsection{$\Lambda_b\to \Lambda_c$ form factors and the Isgur-Wise
function}

\begin{figure}
\begin{center}
\scalebox{0.8}{\includegraphics{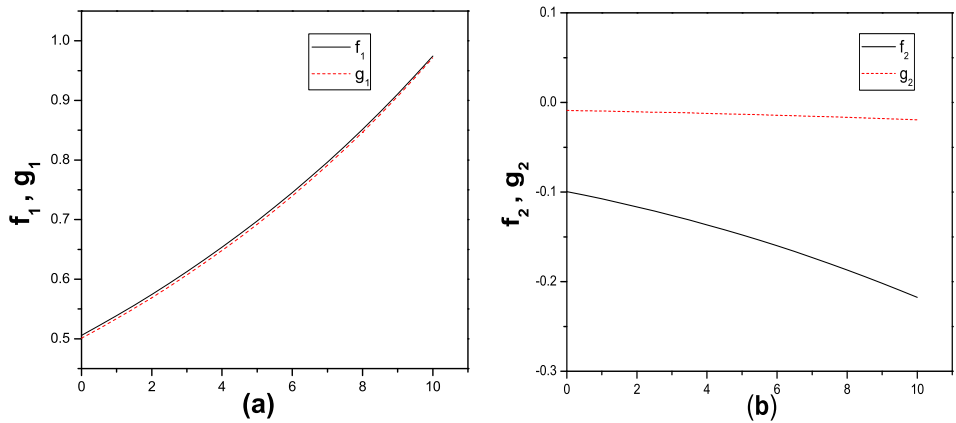}}
\end{center}
\caption{(a) Form factors  $ f_1$ and $g_1$ \,\,\,  (b)
 Form factors  $f_2$ and $g_2$}\label{f2}
\end{figure}

Because the calculation of form factors is performed in the frame
$q^+=0$ with $q^2=-q^2_{\perp}\leq 0$, only the values of the form
factors in the space-like region can be obtained. The advantage of
this choice is that the so-called Z-graph contribution arising
from the non-valence quarks vanishes. In this study, another
advantage is that it simplifies the calculation of baryonic matrix
elements. In order to obtain the physical form factors, an
analytical extension from the space-like region to the time-like
region is required. The form factors in the space-like region can
be parameterized in a three-parameter form as
 \begin{eqnarray}\label{s14}
 F(q^2)=\frac{F(0)}{\left(1-\frac{q^2}{M_{\Lambda_b}^2}\right)
  \left[1-a\left(\frac{q^2}{M_{\Lambda_b}^2}\right)
  +b\left(\frac{q^2}{M_{\Lambda_b}^2}\right)^2\right]},
 \end{eqnarray}
where $F$ represents the form factor $f_{1,2}$ and $g_{1,2}$. The
parameters $a,~b$ and $F(0)$ are fixed by performing a
three-parameter fit to the form factors in the space-like region
which were obtained in previous sections. We then use these
parameters to determine the physical form factors in the time-like
region. The fitted values of $a,~b$ and $F(0)$ for different form
factors $f_{1,2}$ and $g_{1,2}$ are given in Table \ref{Tab:t4}.
The $q^2$ dependence of the form factors is plotted in Fig.
\ref{f2}.

\begin{table}
\caption{The $\Lambda_b\to \Lambda_c$ form factors given in the
  three-parameter form.}\label{Tab:t2}
\begin{ruledtabular}
\begin{tabular}{cccc}
  $F$    &  $F(0)$ &  $a$  &  $b$ \\
  $f_1$  &   0.50568     &   1.00    & 0.75   \\
  $f_2$  &   -0.09943      &   1.50    &  1.43  \\
  $g_1$  &    0.50087     &     1.00  &  0.70  \\
  $g_2$  &      -0.00889   &     1.50  & 1.45
\end{tabular}
\end{ruledtabular}
\end{table}

From Table \ref{Tab:t2} and Fig. \ref{f2}, we find that the form
factors $f_1$ and $g_1$ are nearly equal. At small recoil, i.e.
large $q^2$ region, there is only a tiny difference between the two
functions. Even at the maximal recoil point $q^2=0$, their
difference is less than 3\%. This can be understood by Eq.
(\ref{s9}) where the difference between $f_2$ and $g_2$ is at order
of $\lqcd^2/(\Mb\Mc)$, a next-to-next-to-leading order in the
$1/m_Q$ expansion. The form factor $f_2$ and $g_2$ are small
comparing to $f_1$ and $g_1$. In practice, $g_2$ is approximately
zero, and $f_2$ is about $20-30$\% of $f_1$ and $g_1$. These
conclusions are consistent with the results of \cite{GKLLW}. From
Table \ref{Tab:t2}, the parameter $a$ for various form factors is
close to 1 and the parameter $b$ is small. The results suggest that
the $q^2$-dependence of $f_i$ and $g_i$ approximately exhibits a
dipole behavior $F(q^2)=\frac{F(0)}{(1-q^2/\Mb)^n}$ with $n=2$.

\begin{figure}[hhh]
\begin{center}
\scalebox{0.8}{\includegraphics{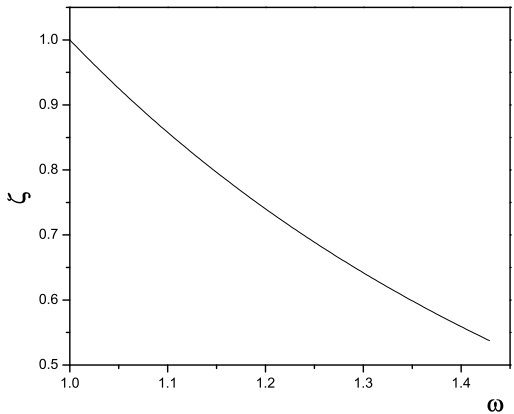}}
\end{center}
\caption{The $\Lambda_{b} \rightarrow \Lambda_{c}$ Isgur-Wise
function $\zeta(\omega)$ with diquark mass $m_{[ud]}=500$ MeV.
}\label{f3}
\end{figure}

In the heavy quark limit, the heavy baryons $\Lambda_b$ and
$\Lambda_c$ have the same scale parameter $\beta^\infty$ in their
wave functions. We choose $\beta^\infty=0.40$ GeV which is equal to
the parameter in the $\Lambda_b$ wave function. The Isgur-Wise
function is usually parameterized by
 \begin{eqnarray}
  \zeta(\omega)=1-\rho^2(\omega-1)+\frac{\sigma^2}{2}(\omega-1)^2+...,
 \end{eqnarray}
where $\rho^2\equiv-\frac{d\zeta(\omega)}{d\omega}|_{\omega=1}$ is
the slope parameter and
$\sigma^2\equiv\frac{d^2\zeta(\omega)}{d\omega^2}|_{\omega=1}$ is
the curvature of the Isgur-Wise function. Our fitted values are
 \beq
 \rho^2&=&1.47, \\
 \sigma^2&=&1.90.
 \eeq
The DELPHI collaboration reported their measurement on the slope
of the Isgur-Wise function $\zeta(\omega)=1-\rho^2(\omega-1)$ as
$\rho^2=2.03\pm 0.46({\rm stat})^{+0.72}_{-1.00}({\rm syst})$ in
the semi-leptonic decay $\Lambda_b\rightarrow \Lambda_c l
\bar{\nu}_l$ \cite{DELPHI}.  The recent theoretical calculations
on the slope parameter $\rho^2$ are: $\rho^2=1.35\pm 0.12$ in QCD
sum rules \cite{HJKL}; $\rho^2=1.51$ in a relativistic quark model
\cite{EFG}.  All the results are in agreement with the experiment
data within the theoretical and experimental errors. The
Isgur-Wise function in the total $\omega$ space is depicted in
Fig. \ref{f3}. The errors in the parameter $\beta^\infty$ has only
a minor effect which is consistent with the B meson case
\cite{LWW}.

\subsection{Semi-leptonic decay of $\Lambda_b \to
\Lambda_c +l\bar{\nu}_l$}

With the form factors and the Isgur-Wise function given in the
above subsection, we are able to calculate the branching ratios
and various asymmetries of $\Lambda_b \to \Lambda_c l\bar{\nu}_l$
decay. Table \ref{Tab:t4} provides our numerical predictions. The
results are presented for two cases: with taking the heavy quark
limit and without taking the heavy quark limit. The ratio of
longitudinal to transverse rates $R>1$ implies that the
longitudinal polarization dominates.

The significant difference for the transverse polarization asymmetry
$a_T$ in the two cases(with or without taking the heavy quark limit)
implies that $a_T$ is sensitive to the heavy quark symmetry breaking
effects. Thus, measurement of $a_T$ may be an ideal probe to test
how well the heavy quark symmetry works in the weak decays of heavy
baryons, not only for the rate estimate, but also other relevant
measurable quantities such as $a_T$. Indeed, for the branching ratio
and the $\Lambda_c$ polarization asymmetry $P_L$, the deviation in
the two cases is at the level of a few percents, thus the heavy
quark limit provides a good approximation.

We also compare our results with the predictions by the relativistic
quark model \cite{EFG}. The two models result in nearly equal
predictions for the longitudinal asymmetry $a_L$ and the
{$\Lambda_c$} polarization asymmetry $P_L$. This confirms the
observation of \cite{CS} that these quantities are less model
dependent.

\begin{table}
\caption{The branching ratios and polarization assymetries of
$\Lambda_b\to \Lambda_c l\bar{\nu}_l$ .}\label{Tab:t4}
\begin{ruledtabular}
\begin{tabular}{c|ccccc}
    &  $Br$ &  $a_L$  &  $a_T$   & $R$    &  $P_L$ \\\hline
  within the heavy quark limit (this work)
    &    $6.2\%$      &   -0.926 & -0.483 & 1.539 & -0.751 \\\hline
  without the heavy quark limit (this work)
    &  $6.3\%$        &   -0.932 & -0.601 & 1.466 & -0.798 \\\hline
  within the heavy quark limit (in \cite{EFG})
    &  $6.2\%$        &  -0.928  & -0.483 & 1.59  & -0.756 \\\hline
  with $1/m_Q$ corrections (in \cite{EFG})
    &   $6.9\%$       &   -0.940 & -0.600 & 1.61  & -0.810
\end{tabular}
\end{ruledtabular}
\end{table}

\subsection{Non-leptonic decays of $\Lambda_b\to\Lambda_c+ M$}

The non-leptonic decays $\Lambda_b\to\Lambda_c +M$ in the
factorization approach have been studied in the previous section.
Now, we present our numerical predictions on the decay rates and
relevant measurable quantities. The CKM matrix elements take
values \cite{PDG06}
 \beq
 && V_{ud}=0.9738, \qquad  V_{us}=0.2257, \qquad V_{cd}=0.230,  \qquad
     \non\\
 && V_{cs}=0.957,~~\qquad  V_{cb}=0.0416, \qquad
 \eeq
and and the effective Wilson coefficient $a_1= 1$. The meson decay
constants are shown in Table \ref{Tab:t5}.

\begin{table}
\caption{Meson decay constants $f$ (in units of
 MeV) \cite{CCH2}.}\label{Tab:t5}
\begin{ruledtabular}
\begin{tabular}{cccccccccc}
  meson & $\pi$ & $\rho$ & $K$ & $K^*$ & $D$ & $D^*$ & $D_s$ & $D_s^*$ & $a_1$ \\\hline
  $f$   & 131   & 216    & 160 & 210   & 200 & 220   & 230   & 230     & 203
\end{tabular}
\end{ruledtabular}
\end{table}

The predictions for the branching ratios and up-down asymmetries are
provided in Table \ref{Tab:t6}. The Tables \ref{Tab:t7} and
\ref{Tab:t8} demonstrate comparisons of our result with that in
other approaches. Some arguments are made in orders:

(1) For the processes with mesons $\pi,\rho,D_s,D_s^*,a_1$ being in
the final states, the corresponding sub-processes are $b\to c\bar u
d$ or $b\to c \bar c s$, which are the Cabibbo-favored processes.
The decay ratios fall in the region $4\times 10^{-3}$ to $1\times
10^{-2}$. They are the dominant decay modes which will be measured
in near future. For the processes with mesons $K,K^*,D,D^*$ in the
final states, the sub-processes are $b\to c\bar u s$ or $b\to c \bar
c d$ which are the Cabibbo-suppressed processes. Their decay ratios
are of order  $(3-5)\times 10^{-4}$.

(2) In the scheme adopted in this work, we obtain the ratio
$\frac{BR(\Lambda_b^0\to\Lambda_c^+ l^-\bar\nu_l)}{BR(\Lambda_b^0
\to\Lambda_c^+\pi^-)}$ to be $16.8$,  and this theoretical
prediction is consistent with the experimental measurement (a
preliminary result) $\frac{BR(\Lambda_b^0\to\Lambda_c^+
l^-\bar\nu_l)}{BR(\Lambda_b^0 \to\Lambda_c^+\pi^-)}=20.0\pm3.0({\rm
stat})\pm 1.2({\rm syst})$ \cite{exp2}.

(3) From Table \ref{Tab:t7}, it is noted that the differences
among the predictions on the branching ratios for non-leptonic
decays by various theoretical approaches are obvious. It is hard
to decide which model is closer to reality at present, because
more precise data are still lacking. It may be more appropriate to
employ the experimental data about the semi-leptonic decay as
inputs to reduce the model dependence of the
$\Lambda_b\to\Lambda_c$ transition form factors as we did in this
work.

(4) All the up-down asymmetries $\alpha$ are negative, this result
reflects the $V-A$ nature of the weak currents.  Table
\ref{Tab:t8} shows that the numerical values of the up-down
asymmetries $\alpha$ predicted by different approaches are nearly
the same except for the process $\Lambda_b^0
\to\Lambda_c^+D_s^{*-}$ where the difference is about 10\%.

\begin{table}
\caption{Branching ratios and up-down asymmetries of non-leptonic
decays $\Lambda_b\to\Lambda_c M$ with the light diquark mass
$m_{[ud]}=500$ MeV.}\label{Tab:t6}
\begin{ruledtabular}
\begin{tabular}{c|cc|cc|}
  & \multicolumn{2}{c|}{within the heavy quark limit~~~~~~}
  & \multicolumn{2}{c|}{without the heavy quark limit~~~~} \\\hline
  & $Br$                & $\alpha$ & $Br$ & $\alpha$          \\\hline
  $\Lambda_b^0\to\Lambda_c^+ \pi^-$  & $4.22\times 10^{-3}$  & $-1$
                        & $3.75\times 10^{-3}$  & $-1$      \\\hline
  $\Lambda_b^0\to\Lambda_c^+ \rho^-$ & $6.07\times 10^{-3}$  & $-0.897$
                        & $6.73\times 10^{-3}$  & $-0.885$  \\\hline
  $\Lambda_b^0\to\Lambda_c^+ K^-$    & $3.41\times 10^{-4}$  & $-1$
                        & $3.02\times 10^{-4}$  & $-1$      \\\hline
  $\Lambda_b^0\to\Lambda_c^+ K^{*-}$  & $3.15\times 10^{-4}$  & $-0.865$
                        & $3.50\times 10^{-4}$  & $-0.857$  \\\hline
  $\Lambda_b^0\to\Lambda_c^+ a_1^-$  & $5.84\times 10^{-3}$  & $-0.758$
                        & $6.49\times 10^{-3}$  & $-0.760$  \\\hline
  $\Lambda_b^0\to\Lambda_c^+ D_s^-$  & $1.18\times 10^{-2}$ & $-0.984$
                        & $1.14\times 10^{-2}$ & $-0.982$  \\\hline
  $\Lambda_b^0\to\Lambda_c^+ D^{*-}_s$& $8.88\times 10^{-3}$ & $-0.419$
                        & $9.96\times 10^{-3}$  & $-0.442$  \\\hline
  $\Lambda_b^0\to\Lambda_c^+ D^-$  & $5.23\times 10^{-4}$ & $-0.987$
                        & $5.01\times 10^{-4}$  & $-0.986$  \\\hline
  $\Lambda_b^0\to\Lambda_c^+ {D^*}^-$& $4.61\times 10^{-4}$& $-0.459$
                        & $5.12\times 10^{-4}$  & $-0.481$
\end{tabular}
\end{ruledtabular}
\end{table}

\begin{table}
\caption{Branching ratios for non-leptonic decays
$\Lambda_b\to\Lambda_c +M$ within different theoretical approaches
(in units of $10^{-2}$).}\label{Tab:t7}
\begin{ruledtabular}
\begin{tabular}{ccccccc}
  & This work & \cite{Cheng} & \cite{MGKIIO} & \cite{FR} & \cite{GMM} & \cite{LLS} \\\hline
  $\Lambda_b^0\to\Lambda_c^+ \pi^-$  & 0.375 & 0.38 & 0.175 & -    & 0.391 & 0.503 \\\hline
  $\Lambda_b^0\to\Lambda_c^+ \rho^-$ & 0.673 & 0.54 & 0.491 & -    & 1.082 & 0.723 \\\hline
  $\Lambda_b^0\to\Lambda_c^+ K^-$    & 0.030 &  -   & 0.013 &      & -     & 0.037 \\\hline
  $\Lambda_b^0\to\Lambda_c^+ K^{*-}$ & 0.035 &  -   & 0.027 & -    & -     & 0.037 \\\hline
  $\Lambda_b^0\to\Lambda_c^+ a_1^-$  & 0.649 &  -   & 0.532 & -    & -     & -     \\\hline
  $\Lambda_b^0\to\Lambda_c^+ D_s^-$  & 1.140 & 1.1  & 0.770 & 2.23 & 1.291 & -     \\\hline
  $\Lambda_b^0\to\Lambda_c^+D^{*-}_s$& 0.996 & 0.91 & 1.414 & 3.26 & 1.983 & -     \\\hline
  $\Lambda_b^0\to\Lambda_c^+ D^-$    & 0.050 &  -   & 0.030 & -     & -     \\\hline
  $\Lambda_b^0\to\Lambda_c^+ {D^*}^-$& 0.051 &  -   & 0.049 & -     & -
\end{tabular}
\end{ruledtabular}
\end{table}

\begin{table}[!h]
\caption{Up-down asymmetries for non-leptonic decays
$\Lambda_b\to\Lambda_c M$ within different theoretical
approaches.}\label{Tab:t8}
\begin{ruledtabular}
\begin{tabular}{cccccc}
  & This work     & \cite{Cheng} & \cite{MGKIIO}  & \cite{FR}    & \cite{LLS}    \\\hline
  $\Lambda_b^0\to\Lambda_c^+ \pi^-$  & -1     & -0.99 & -0.999 & -     &  -1     \\\hline
  $\Lambda_b^0\to\Lambda_c^+ \rho^-$ & -0.885 & -0.88 & -0.897 & -     &  -0.885 \\\hline
  $\Lambda_b^0\to\Lambda_c^+ K^-$    & -1     &  -    & -1     &       &  -1     \\\hline
  $\Lambda_b^0\to\Lambda_c^+ K^{*-}$ & -0.857 &  -    & -0.865 & -     &  0.885  \\\hline
  $\Lambda_b^0\to\Lambda_c^+ a_1^-$  & -0.760 &  -    & -0.758 & -     &  -      \\\hline
  $\Lambda_b^0\to\Lambda_c^+ D_s^-$  & -0.982 & -0.99 & -0.984 & -0.98 &  -      \\\hline
  $\Lambda_b^0\to\Lambda_c^+D^{*-}_s$& -0.442 & -0.36 & -0.419 & -0.40 &  -      \\\hline
  $\Lambda_b^0\to\Lambda_c^+ D^-$    & -0.986 &  -    & -0.987 & -     &  -      \\\hline
  $\Lambda_b^0\to\Lambda_c^+ {D^*}^-$& -0.481 &  -    & -0.459 & -     &  -
\end{tabular}
\end{ruledtabular}
\end{table}

\section{Conclusions}

In this work, we investigate extensively the $\Lambda_b\to\Lambda_c$
transition form factors in the light-front approach and make
predictions on the rates for the semi-leptonic decay
$\Lambda_b\to\Lambda_c l\bar\nu_l$ and non-leptonic two-body decays
$\Lambda_b\to\Lambda_c+ M$. In the light-front quark model, we adopt
the diquark picture for the heavy baryons. It is believed that for
the the heavy baryons which contain at least one heavy quark, the
quark-diquark picture seems to work well, therefore one can employ
it for evaluating the hadronic matrix elements of
$\Lambda_b\to\Lambda_c$ transitions which are dominated by
non-perturbative QCD effects. The light scalar diquark mass
determined from the data on the semi-leptonic decay is about $500$
MeV. Our numerical results show that the $q^2$-dependence of the
momentum transfer of different form factors has a dipole-like
behavior. The slope parameter of the universal Isgur-Wise function
is found to be consistent with that obtained by fitting experimental
data. The small difference for the branching ratio of the
semi-leptonic decay with and without the heavy quark limit implies
that the heavy quark symmetry is good in the heavy bottom baryon
system. However on the other aspect, the transverse polarization
asymmetry is shown to be sensitive to the heavy quark symmetry
breaking, and it is worth further and more accurate studies. Our
results for the exclusive non-leptonic two-body decays
$\Lambda_b\to\Lambda_c+ M$ is modest among the predictions by other
approaches. The semi-leptonic to non-leptonic $\Lambda_c^+\pi^-$
decay ratio is well in accord with the experimental measurements.
The non-leptonic decays, so far have not been accurately measured,
and there are only upper bounds for some channels, so that it is
still hard to judge the closeness of the present models to the
physical reality yet. Fortunately the LHCb will run and a remarkable
amount of data on $\Lambda_b$ production and decay will be
accumulated in the future LHCb, then one may be able to verify the
different models.

\section*{Acknowledgement}

This work was supported in part by NNSFC under contract Nos.
10475042, 10745002 and 10705015.


\begin{thebibliography}{99}

\bibitem{DELPHI} J. Abdallah $et$ $al$., DELPHI Collaboration, Phys. Lett. B {\bf 585},
 63-84 (2004).

\bibitem{AHN} C. Albertus, E. Hernandez and J. Nieves, Phys. Rev. D {\bf 71}, 014012 (2005).

\bibitem{SLL}H. Shih, S. Lee and H. Li, Phys. Rev. D {\bf 61}, 114002 (2000).

\bibitem{GKLLW} P. Guo, H. Ke, X. Li, C. Lu and Y. Wang, Phys. Rev. D {\bf 75},
 054017 (2007).

\bibitem{HJKL}M. Huang, H. Jin, J. K$\rm{\ddot{o}}$rner and C. Liu, Phys.
 Lett. B {\bf 629}, 27 (2005).

\bibitem{PWC} M. Pervin, W. Roberts and S. Capstick, Phys. Rev. C {\bf 72}, 035201 (2005).

\bibitem{EFG} D. Ebert, R.  Faustov and V. Galkin, Phys. Rev.
 D {\bf 73}, 094002 (2006).

\bibitem{PDG06} W. Yao $et$ $al$., Partical Data Group, J. Phys. G
 {\bf 33}, 1 (2006).

\bibitem{Factorization} A. Ali, G. Kramer, C.D. Lu, \prd {\bf 58}, 094009
 (1998); Y. Chen, H. Cheng, B. Tseng, K. Yang, \prd {\bf 60}, 094014
 (1999); C. Chen, C. Geng, Z. Wei, Eur. Phys. J. C {\bf 46}, 367 (2006).

\bibitem{CS} F. Cardarelli and S. Simula, Phys. Rev. D {\bf 60},
 074018 (1999).

\bibitem{light} M. Terent'ev, Sov. J. Nucl. Phys. {\bf 24}, 106 (1976);
 V. Berestetsky and M. Terent'ev, $ibid$. {\bf 24}, 547 (1976); {\bf 25}, 347 (1977);
 P. Chung, F. Coester, and W. Polyzou, Phys. Lett. B {\bf 205}, 545 (1988).

\bibitem{Jaus} W. Jaus, Phys. Rev.  D {\bf 41}, 3394 (1990);
  D {\bf 44}, 2851 (1991); {\bf 60}, 054026 (1999).

\bibitem{meson2}C. Ji, P. Chung and S. Cotanch, Phys. Rev. D \textbf{45}, 4214
 (1992).

\bibitem{CCH1} H. Cheng, C. Cheung and C. Hwang, Phys.
  Rev. D {\bf 55}, 1559 (1997).

\bibitem{CCH2} H. Cheng, C. Chua and C. Hwang, Phys.
  Rev. D {\bf 69}, 074025 (2004).

\bibitem{HW} C. Hwang and Z. Wei, J. Phys. G {\bf 34}, 687
 (2007).

\bibitem{DJS} H. Dosch, M. Jamin and B. Stech, Z. Phys. C {\bf 42}, 167 (1989).

\bibitem{kroll}P. Kroll, B. Quadder and W. Schweiger, Nucl. Phys. {\bf B 316},373
 (1989); P. Ball, H.G. Dosch,  Z. Phys. {\bf C 51}, 445 1991;
 J. K$\rm{\ddot{o}}$rner and P. Kroll, Phys. Lett. {\bf B 293}, 201 1992;
 R. Jakob, P. Kroll, M. Schurmann and  W. Schweiger, Z.
 Phys. {\bf A 347}, 109 (1993); J. Bolz, P. Kroll and J.
 K$\rm{\ddot{o}}$rner, Z. Phys. {\bf A 350}, 145 (1994).

\bibitem{wilczek} F. Wilczek, arXiv: hep-ph/0409168.

\bibitem{yu} Y. Yu, H. Ke, Y. Ding, X. Guo, H. Jin, X. Li, P. Shen
 and G. Wang, Commun. Theor. Phys. {\bf 46}, 1031  (2006); Y. Yu, H.
 Ke, Y. Ding, X. Guo, H. Jin, X. Li, P. Shen and G. Wang, arXiv:
 hep-ph/0611160.

\bibitem{MQS} B. Ma, D. Qing and I. Schmidt, Phys. Rev. C {\bf
 66},  048201 (2002).


\bibitem{pentaquark2} H. Cheng and C. Chua , Phys. Rev. D {\bf 70},
 034007 (2004).

\bibitem{BD} S. Brodsky and S. Drell, Phys. Rev. D {\bf 22}, 2236
 (1980).

\bibitem{Schlumpf} F. Schlumpf, Phys. Rev. D {\bf 47}, 4114
 (1993); {\bf 49}, 6246(E) (1994).

\bibitem{HQS} For a review, see M. Neubert, Phys. Rept. {\bf 245},
  259 (1994).

\bibitem{IW} N. Isgur and M. Wise, Nucl. Phys. B {\bf  348}, 276 (1991);
 H. Georgi, Nucl. Phys. B {\bf 348}, 293 (1991).

\bibitem{CZL} C. Cheung, W. Zhang and G. Lin,
  Phys. Rev. D {\bf 52}, 2915 (1995).

\bibitem{KKP}
J.  K$\rm\ddot{o}$rner and M. Kr$\rm\ddot{a}$mer, Phys. Lett. B
{\bf275}, 495 (1992); P. Bialas, J.  K$\rm\ddot{o}$rner, M.
Kr$\rm\ddot{a}$mer, and K. Zalewski, Z. Phys. C {\bf57}, {115}
(1993); J. K$\rm{\ddot{o}}$rner, M. Kr$\rm\ddot{a}$mer and D.
Pirjol,
 Prog. Part. Nucl. Phys. {\bf 33}, 787 (1994).

\bibitem{KK} J. K$\rm{\ddot{o}}$rner and M. Kr$\rm\ddot{a}$mer, Z. Phys. C {\bf 55}, 659
 (1992).

\bibitem{Cheng} H. Cheng, Phys. Rev. D {\bf 56}, 2799 (1997).

\bibitem{Guo} X. Guo and T. Muta, Phys. Rev. {\bf D 54}, 4629 (1996).

\bibitem{HKLW} M. Heb, M. Karsch, E. Laermann and I. Wetzorke,
 arXiv: hep-lat/9804023.


\bibitem{LWW} C. Lu, W. Wang and Z. Wei, Phys. Rev. {\bf D 76}, 014013 (2007).

\bibitem{MGKIIO} R. Mohanta, A. Giri, M. Khanna, M. Ishida, S. Ishida
 and M. Oda, Prog. Theor. Phys. {\bf 101}, 959(1999).

\bibitem{FR} Fayyazuddin and Riazuddin, Phys. Rev. D {\bf 58}, 014016 (1998).

\bibitem{GMM} A.  Giri, L. Maharana and R. Mohanta, Mod. Phys.
 Lett. A {\bf 13}, 23 (1998).

\bibitem{LLS} J. Lee, C. Liu and H. Song, Phys. Rev. D {\bf
 58}, 014013 (1998).

\bibitem{exp2} S. Yu, arXiv: hep-ex/0504059.





\end{thebibliography}
\end{document}